\documentclass[twocolumn]{aastex61}
\pdfoutput=1
\usepackage[latin1]{inputenc}
\usepackage{amsmath}
\usepackage{amsfonts}
\usepackage{amssymb}
\usepackage{chngcntr}

\usepackage{graphicx}
\usepackage[varg]{txfonts}
\usepackage{mwe} 
\usepackage{url}
\usepackage{longtable}
\usepackage{float}

\shorttitle{Radiative efficiency and Reionization Contribution}
\shortauthors{Ananna et al.}

\begin{document}
\title{Accretion History of AGN III: Radiative Efficiency and AGN Contribution to Reionization}
	
\author{Tonima Tasnim Ananna}
\affiliation{Department of Physics \& Astronomy, Dartmouth College, 6127 Wilder Laboratory, Hanover, NH 03755, USA}
\affiliation{Department of Physics, Yale University, PO BOX 201820, New Haven, CT 06520-8120}
\affiliation{Yale Center for Astronomy and Astrophysics, P.O. Box 208120, New Haven, CT 06520, USA 0000-0002-5554-8896}
\email{Tonima.Ananna@dartmouth.edu}

\author{C. Megan Urry}
\affiliation{Department of Physics, Yale University, PO BOX 201820, New Haven, CT 06520-8120}
\affiliation{Yale Center for Astronomy and Astrophysics, P.O. Box 208120, New Haven, CT 06520, USA 0000-0002-5554-8896}
\affiliation{Department of Astronomy, Yale University, P.O. Box 208101, New Haven, CT 06520, USA}

\author{Ezequiel Treister}
\affiliation{Instituto de Astrof\'{\i}sica, Facultad de F\'{\i}sica, Pontificia Universidad Cat\'{o}lica de Chile, Casilla 306, Santiago 22, Chile, 0000-0001-7568-6412}

\author{Ryan C. Hickox}
\affiliation{Department of Physics \& Astronomy, Dartmouth College, 6127 Wilder Laboratory, Hanover, NH 03755, USA}

\author{Francesco Shankar}
\affiliation{Department of Physics and Astronomy, University of Southampton, Highfield, SO17 1BJ, UK}

\author{Claudio Ricci}
\affiliation{N\'ucleo de Astronom\'ia de la Facultad de Ingenier\'ia, Universidad Diego Portales, Av. Ej\'ercito Libertador 441, Santiago, Chile}
\affiliation{Kavli Institute for Astronomy and Astrophysics, Peking University, Beijing 100871, China}

\author{Nico Cappelluti}
\affiliation{Department of Physics, University of Miami, Coral Gables, FL 33124, U.S.A.}

\author{Stefano Marchesi}
\affiliation{INAF - Osservatorio di Astrofisica e Scienza dello Spazio di Bologna, Via Piero Gobetti, 93/3, 40129, Bologna, Italy}
\affiliation{Department of Physics and Astronomy, Clemson University, Kinard Lab of Physics, Clemson, SC 29634, USA}

\author{Tracey Jane Turner}
\affiliation{Eureka Scientific, 2452 Delmer Street Suite 100, Oakland, CA 94602-
3017, USA}

\begin{abstract}
{ The cosmic history of supermassive black hole (SMBH) growth is important for understanding galaxy evolution, reionization and the physics of accretion. Recent \textit{NuSTAR}, \textit{Swift}-BAT and \textit{Chandra} hard X-ray surveys have provided new constraints on the space density of heavily obscured Active Galactic Nuclei (AGN). 
Using the new X-ray luminosity function derived from these data, we here estimate the accretion efficiency of SMBHs and their contribution to reionization. 
We calculate the total ionizing radiation from active galactic nuclei (AGN) as a function of redshift, based on the X radiation and distribution of obscuring column density, converted to UV wavelengths. 
Limiting the luminosity function to unobscured AGN only, our results agree with current UV luminosity functions of unobscured AGN.
For realistic assumptions about the escape fraction, the contribution of all AGN to cosmic reionization is $\sim4$ times lower than the galaxy contribution (23\% at $z\sim6$).
Our results also offer an observationally constrained prescription that can be used in simulations or models of galaxy evolution.
To estimate the average efficiency with which supermassive black holes convert mass to light, we compare the total radiated energy, converted from X-ray light using a bolometric correction, to the most recent local black hole mass density. The most likely value, $\eta \sim 0.3-0.34$, approaches the theoretical limit for a maximally rotating Kerr black hole, $\eta=0.42$, implying that on average growing supermassive black holes are spinning rapidly.}

\end{abstract}


\section{Introduction}\label{sec:intro}

Understanding the cosmological growth of black holes in active galactic nuclei (AGN) is important for understanding the co-evolution of supermassive black holes and galaxies (see \citealp{alexanderhickox2012} for a review) because
modern theories of galaxy evolution usually invoke energy injection by AGN (e.g., \citealp{genel2014,sville2015}).
An accurate history of AGN energy deposition can reduce uncertainties in those models. 
AGN radiation also makes a potentially significant contribution to reionization in the early universe.
Furthermore, comparing the total light emitted by AGN to the mass accreted over time yields an estimate of the accretion efficiency of black holes
\citep{soltan1982,smallblandford1992}. This radiative efficiency offers insight into the physics close to the black hole since, according to standard accretion disk theory, black hole spin determines the radius of the innermost stable circular orbit \citep{pagethorne1974}.

These important issues relate to the total light emitted by AGN, which is derived from the space density of AGN as a function of redshift, luminosity and obscuration, along with an appropriate bolometric correction. An unbiased measurement of the space density of AGN requires accounting for heavily obscured AGN. Many lines of evidence suggest that AGN undergo a prolonged period of obscured growth---for example, this is required to reproduce the shape of the cosmic X-ray ``background'' spectrum (e.g., \citealp{comastri1995,gilli2001,tresiter2004,ananna2020}). While optical and ultraviolet (UV) light are heavily absorbed by dust,
hard X-rays can escape even heavily obscured AGN; X-ray surveys are also highly efficient \citep{HandA2018} in detecting AGN because X-ray luminosities above $L_{\rm 2-10~keV}>10^{42}$~erg/s are easily distinguished from star forming contaminants. At high energies (E $>$ 10 keV), in particular, X-ray observatories such as \textit{NuSTAR} \citep{harrison2013} and the \textit{Neil Gehrels Swift Observatory's} Burst Alert Telescope (BAT; \citealp{barthelmy2005}) produced a largely unbiased census of black hole growth \citep{harrison2016}. Many previous works have estimated the intrinsic space density of AGN
based on soft X-ray data ($<$ 3 keV; \citealp{maccacaro1991,boyle1993,comastri1995,jones1997,page1997,miyaji2000,gilli2001}). Higher energy X-ray surveys 
yield higher space densities of obscured objects 
(\citealp{boyle1998,cowie2003,ueda2003,gilli2007,treister2009,treister2010,ueda2014,aird2015,buchner2015}); however, these XLFs were constrained before the newest data from \textit{NuSTAR} and \textit{Swift}-BAT 70-month survey became available. 
The sensitivity of \textit{NuSTAR} and \textit{Swift}-BAT at energies $E>10$~keV gives a much more unbiased view of the AGN population in terms of obscuration, and incorporating these new data gives us an unprecedented opportunity to construct an unbiased census of SMBH growth.

In \citeauthor{ananna2019} (\citeyear{ananna2019}, henceforth A19), we constrained the space densities of AGN up to redshift $z\sim5$ over the luminosity range $L_{\rm 2-10~keV} = 10^{41}$ erg/s $- 10^{47}$ erg/s, at obscuration levels (quantified by equivalent hydrogen column density) 
log (${\rm N}_{\rm H}$/cm$^{-2}$) $=20-26$, using X-ray data from \textit{Swift}-BAT, \textit{NuSTAR}, \textit{XMM}-Newton \citep{jansen2001} and \textit{Chandra} \citep{weisskopf2002}. 
We were able to satisfy simultaneously more than a dozen new observed constraints (summarized in Table~1 of A19) that previous X-ray luminosity functions (XLFs) could not. 
Compared to previous works, the A19 XLF
includes a larger fraction of obscured objects, with Compton-thick objects comprising roughly half the AGN population.

In this paper we use the A19 XLF to quantify the radiative history of AGN and the spin and radiative efficiency of accreting supermassive black holes,
taking into account the dispersion in X-ray-to-bolometric correction by marginalizing over the uncertainty in conversion, and using the most up-to-date local black hole mass density \citep{shankar2019nature}.

We also use the A19 XLF to explore the contribution of AGN to the cosmic reionization budget. As the number of faint AGN at high redshifts is uncertain, due both to intrinsically low luminosities and to obscuration, this quantity has been difficult to establish. \citet{kashikawa2015} and \citet{onoue2017} evaluated the faint end of the luminosity function using a handful of rest-frame optical/UV-selected quasars at $z\sim6$, concluding that 
they contribute 1$-$12\% of the ionizing photon budget at 
that redshift even
if the escape fraction of photons into the intergalactic medium (IGM) is unity. \citet{shankarmathur2007} and \citet{fricci2017} also found a similar contribution to reionization by AGN using X-ray surveys. In contrast, \citet{giallongo2015}, 
based on their faint-end UV luminosity function at $4< z <6.5$, 
concluded that AGN alone could produce enough ionizing photons. Previous works, such as \citet{fricci2017}, considered the contribution to reionization from X-ray detected objects with $\log~(N_H/{\rm cm^2}) < 22$. It is possible that Compton-thin and Compton-thick objects could also contribute significantly to reionization in a purely unified scheme, wherein on average all AGN have the same opening angle and therefore a constant escape fraction. It is also possible that at least some obscured AGN reside in different environments and have different escape fractions. In this work, we consider both scenarios. 

Another source of uncertainty in using an XLF to estimate the AGN contribution to reionization is the scatter in the X-ray-to-UV conversion. To address this issue, we marginalize over the dispersion in the conversion factor.


While previous works have computed the reionization contribution (e.g., \citealp{fricci2017}) or calculated the radiative efficiency (e.g., 
\citealp{shankar2019nature}), 
they were based on XLFs that 
fail to match the new \textit{NuSTAR} and \textit{Swift}-BAT results, as well as the Compton-thick constraints from deep \textit{Chandra}-COSMOS data \citep{civano2015,lanzuisi2018}. 
The A19 XLF incorporates these new constraints and thus provides new information about the total radiation from AGN.

Specifically, Compton-thick number densities have been quite uncertain over the last decades, in part because of insufficient high energy X-ray data. 
For example, the \citet{ueda2014} sample lacked enough Compton-thick objects to constrain their number densities, so they simply assumed that the number densities of Compton-thin and Compton-thick objects were equal. \citet{shankar2019nature} explored the effect of doubling the Compton-thick space densities for U14; however, much higher space densities of Compton-thick objects are required to fit the newest constraints (see Figure~10 of A19). 
Similarly, the faint AGN number counts from lower energy bands, such as very deep Chandra fields 
disagree with previous luminosity functions (see Fig.~12 of A19), as do the overall number counts and $N_H$ distributions for unabsorbed and Compton-thin objects from \textit{Swift}-BAT (see Fig.~10 in A19; previous luminosity functions fail above $\log N_H > 21$~atoms/cm$^2$).
Only the A19 XLF fits the full sweep of data now available.
The present work explores what this new, highly constrained XLF means for radiative efficiency and AGN contribution to reionization. 

Our method and findings are structured as follows: \S~\ref{sec:method} describes 
how we derive accretion history, efficiency and reionization from the XLF;
in \S~\ref{sec:results} we report our results; and in \S~\ref{sec:conclusion} we discuss these results. We adopt a
flat cosmology with $\Omega_{m} = $ 0.3, $\Omega_{\Lambda} = $ 0.7 and h $=$ 0.7.

\section{Method}\label{sec:method}

The A19 XLF, which fits all currently available X-ray constraints, 
is used in this work to estimate the radiative efficiency of black hole accretion and calculate the overall AGN contribution to reionization. 
In this section, we briefly describe 
how radiative efficiency and 
cosmic reionization are calculated from the XLF.

\begin{figure}[t]
	\centering
	\includegraphics[width=1.\linewidth]{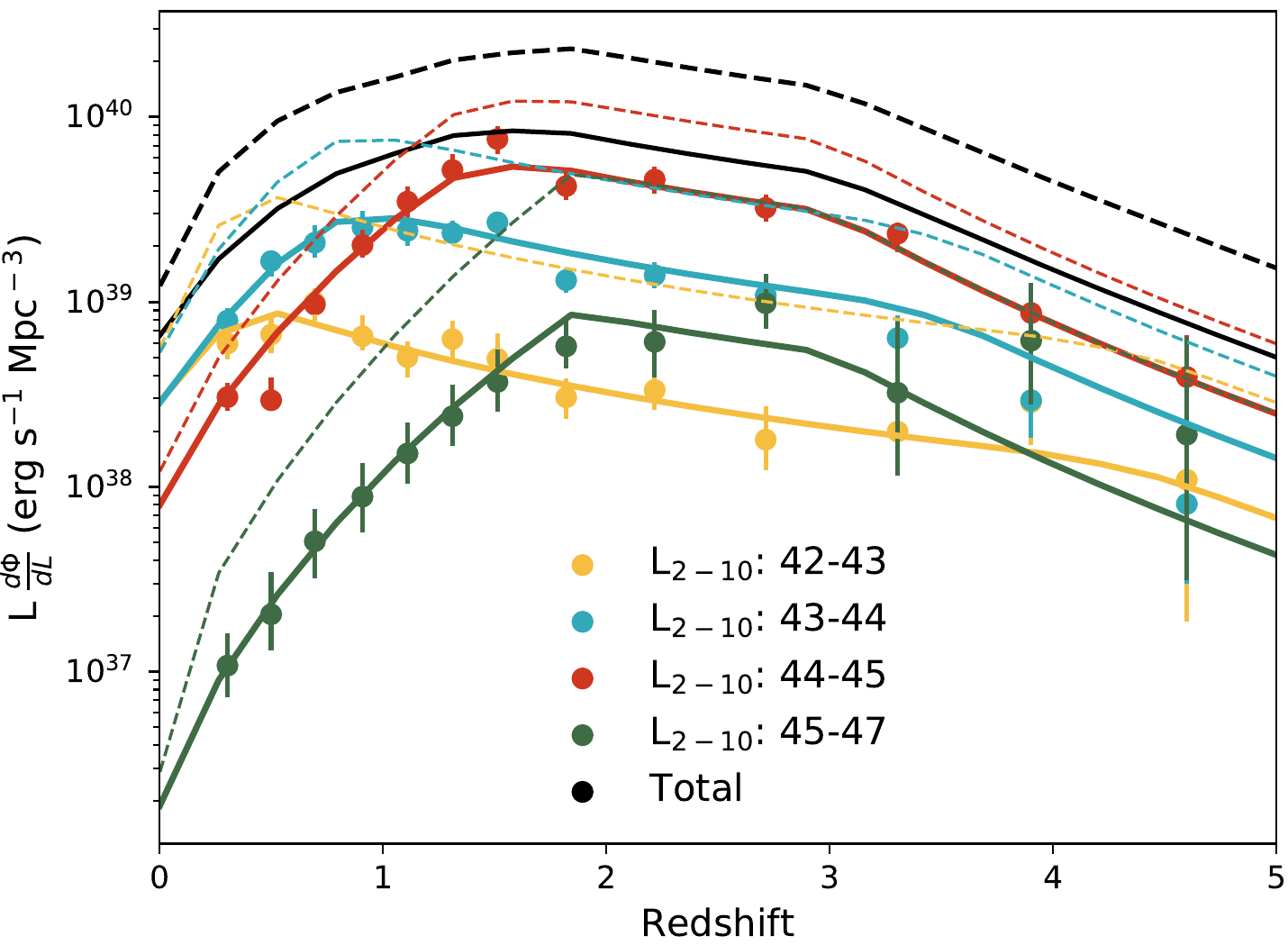}
	\caption{Luminosity density as a function of redshift, calculated from the A19 luminosity function. 
	The light from AGN with $N_{\rm H} < 10^{24}$~atoms cm$^{-2}$ (\textit{solid lines}) matches well the observed data
	(from Figure~12 of \citealp{ueda2014}).
	Compton-thick AGN ($N_{\rm H} > 10^{24}$~atoms cm$^{-2}$) represent a comparable contribution to the total light (included in the \textit{dashed lines}) but have been generally missed by rest-frame UV or soft X-ray surveys.
	Black lines represent total AGN light integrated over all luminosities, while colored lines correspond to luminosity bins denoted in the label.}
	\label{fig:A19_lphi} 
\end{figure}

\subsection{Cosmic History of Supermassive Black Hole Growth}

As supermassive black holes accrete material, they emit radiation across most of the electromagnetic spectrum. The A19 XLF gives the space density of AGN as a function of 2-10~keV luminosity, redshift, and obscuring column density, $\Phi (L_{\rm X},z,N_{\rm H})$. Integrating over the latter for $N_{\rm H}<10^{24}$~atoms cm$^{-2}$, and over various ranges in luminosity, we obtain X-ray luminosity density as a function of redshift (solid lines in Fig.~\ref{fig:A19_lphi}). The colored lines predicted by the A19 XLF, which fit the data well, show that peak luminosity density occurs at higher redshifts for higher luminosities, as found previously by \citet{hasinger2005}, \citet{hasinger2008}, \citet{ueda2014}, and \citet{miyaji2015}. Integrating over all $N_{\rm H}$ adds Compton-thick objects, which are mostly missed by rest-frame UV or soft X-ray surveys, roughly doubling the emitted radiation (dashed lines in Fig.~\ref{fig:A19_lphi}).

The light emitted is connected to the rate of mass accreted onto the black hole via the relationship $L = (\eta + \eta_{\rm kin}) \dot{M}_{\rm acc}c^2$, where $\eta$ is the radiative efficiency of the supermassive black hole (SMBH), $\eta_{\rm kin}$ corresponds to the accretion energy that goes into relativistic jets or fast winds \citep{shankar2008,merloni2007,lafranca2010}, $L$ is the bolometric luminosity, and $\dot{M}_{\rm acc}$ is the mass accretion rate.
\citet{soltan1982} was the first to estimate the radiative efficiency of black hole accretion in this way (without the kinetic term). We note that mergers can also increase SMBH mass, meaning that the summed mass of all SMBH seeds should properly be subtracted from the local black hole mass density before estimating accretion efficiency; however, current estimates suggest this fixed term is a small fraction of the local SMBH mass density \citep{volonteri2005,berti2008,shankar2009,shankar2010,li2012,shankar2013,aversa2015}, so we ignore it here.

Theoretically, if the angular momentum and spin of a rapidly rotating system are aligned, 
the radiative efficiency can reach $\eta \simeq 0.42$, whereas the theoretical limit for non-rotating black holes is 0.057 \citep{thorne1974}. 
Following the Soltan argument,
previous works 
have produced estimates of the average radiative efficiency over cosmic time in the range $\eta\sim 0.05-0.1$
(e.g., \citealp{chokshiturner1992,smallblandford1992,yutremaine2002,marconi2004,shankar2004,caoli2008,cao2010,li2012,ueda2014}), 
depending on the AGN luminosity function and black hole mass density assumed. These low values were interpreted as an indication that maximally-rotating Kerr black holes do not dominate the population.

As per the Soltan argument, the mass density of SMBH at a certain redshift is the result of accretion over earlier cosmic times. Specifically, for a bolometric AGN luminosity function $\Phi (L, z)$, where $L$ is the bolometric luminosity, the accumulated mass density is given by: 

\begin{equation}\label{eqn:rho}
\rho (z) = \int^{z_{S}}_{z} \int^{L_{max}}_{L_{min}} \frac{(1 - \eta - \eta_{\rm kin} )~L}{\eta~c^2} 
\Phi
(L, z) d\log L \frac{dt}{dz} dz ,
\end{equation}
%

\noindent
where $z_S$ is the upper redshift limit at which substantial black hole growth begins. 
The A19 luminosity function is constrained using data up to a redshift of $z\sim5$ but we extrapolate the space densities to redshift $z_S=6$, and 
integrate over bolometric luminosities 10$^{42}-10^{50}$ erg/s. We note that the narrower integration interval in luminosity (i.e., 10$^{42}-10^{48}$ erg/s) used by \citet{shankar2019nature} leads to a slightly lower radiative efficiency than the one we report below. We converted X-ray luminosity to bolometric using the most updated luminosity-dependent correction from \citet{duras2020}. 

In previous works (e.g., \citealp{shankar2019nature}), $\phi (\log L_{bol}, z)\\ d \log L_{bol}$ was calculated as follows:
\begin{equation}
    \phi (\log L_{bol}, z) d \log L_{bol} = \phi (\log L_{X}, z) \frac{d \log L_{X}}{d \log L_{bol}} d \log L_{bol} ,
\end{equation}
where the derivative, $\frac{d \log L_{X}}{d \log L_{bol}}$, depends on the relations, $L_{bol} = K_{X} (L_{X})~L_{X}$ or $L_{bol} = K_{\rm bol} (L_{bol})~L_{X}$ (see \citealp{marconi2004}, \citealp{hopkins2007} and \citealp{duras2020} for examples). To take into account the dispersion in the $L_{bol}$ to $L_{X}$ conversion, we used the approach outlined by \citet{georgakylas2010}:

\begin{multline}
    \Phi (\log L_{bol}) = \int \Phi (\log L_{\rm 2-10keV}) \left(\frac{1}{\sigma_{\log L_{\rm bol}} \sqrt{2 \pi}} \right) \\
    \exp{\frac{-[\log L_{bol}-\langle \log L_{\rm bol} (L_{2-10keV}) \rangle ]^2}{2 \sigma_{\log L_{\rm bol}}^2}}~dL_{\rm 2-10keV} .
\end{multline}

\noindent
\noindent
We use the bolometric conversion presented in \citet{duras2020} as the average value for X-ray to bolometric conversion, and estimate an empirical dispersion, $\sigma_{\log L_{\rm bol}} = 0.2$, from Figure~4 of \citet{duras2020}.

Assuming no dependence of radiative efficiency on redshift, mass, or luminosity, the right-hand side of Equation~\ref{eqn:rho} varies linearly with $\frac{(1 - \eta - \eta_{\rm kin} )}{\eta~c^2}$. Some authors have reported $\eta_{\rm kin} \simeq 0.1$ \citep{merloni2007,shankar2008,lafranca2010,ghisellini2013}, while \citet{zubrovas2018} suggests it might be higher. Others have reported much lower kinetic efficiency, $\eta_{\rm kin} = 0.03$ \citep{gaspari2017}. Following \citet{shankar2019nature},
we considered two values, 0 and 0.15, that bracket this range in order to explore how kinetic efficiency might affect the average radiative efficiency for a given local black hole mass density.

To evaluate the confidence interval in $\eta$, we used a Bayesian approach, 
carrying out Monte Carlo Markov Chain (MCMC) sampling using an ensemble sampler with 200 walkers. In this sampler, we calculated $\rho (z=0)$ from Equation~\ref{eqn:rho} assuming the A19 XLF. (Basing the luminosity function on an X-ray-selected sample rather than on optical- or UV-selected sample ensures a more complete representation of the AGN population.) Each walker initiates at a random value of $\eta$ and tries to minimize the difference between the model prediction of $\rho (z=0)$ and the observed mass density by changing $\eta$; together the ensemble converges on a distribution of viable values for efficiency. We repeated these calculations using the \citeauthor{ueda2014} (\citeyear{ueda2014}) 
XLF and compare the results below.

Ideally we would compare SMBH mass density to AGN XLF as a function of redshift; however, since the mass 
functions
are uncertain at higher redshifts, we only compare integrated results locally. \citet{li2011} and \citet{vika2009} reported a local SMBH mass density of $\rho_{\rm SMBH} = 4.5-5 \times 10^5 M_\odot$ Mpc$^{-3}$.
\citet{shankar2016,shankar2017,shankar2019nature,shankar2020mnras} showed that selection effects can artificially increase the normalization of the black hole mass-velocity dispersion (M$_{\rm BH}-\sigma$) relation because large black holes are easier to detect and high masses and velocities are easier to measure; 
their estimate was $\rho_{\rm SMBH} = 1.2 \times 10^5 M_\odot$ Mpc$^{-3}$. 

We also convert radiative efficiency to black hole spin according to the highly non-linear relation in Figure~6 of \citet{reynolds2012}. Zero spin is associated with low radiative efficiency, $\eta=0.057$; efficiency increases slowly to $\eta \sim 0.1$ for a spin of roughly $\sim 0.8$; and for higher spin values, efficiency increase much more rapidly. 
In \S\S~3 and 4, we report the radiative efficiency and spin for two values of local SMBH mass density.

\subsection{Calculating Contribution to Cosmic Reionization}
\label{ssec:UVconv}

\begin{figure*}[t]
	\centering
	\begin{picture}(50,1)
	\put(-230,-150){\rotatebox{90}{Probability density}}
	\end{picture}
    \includegraphics[width=0.45\linewidth]{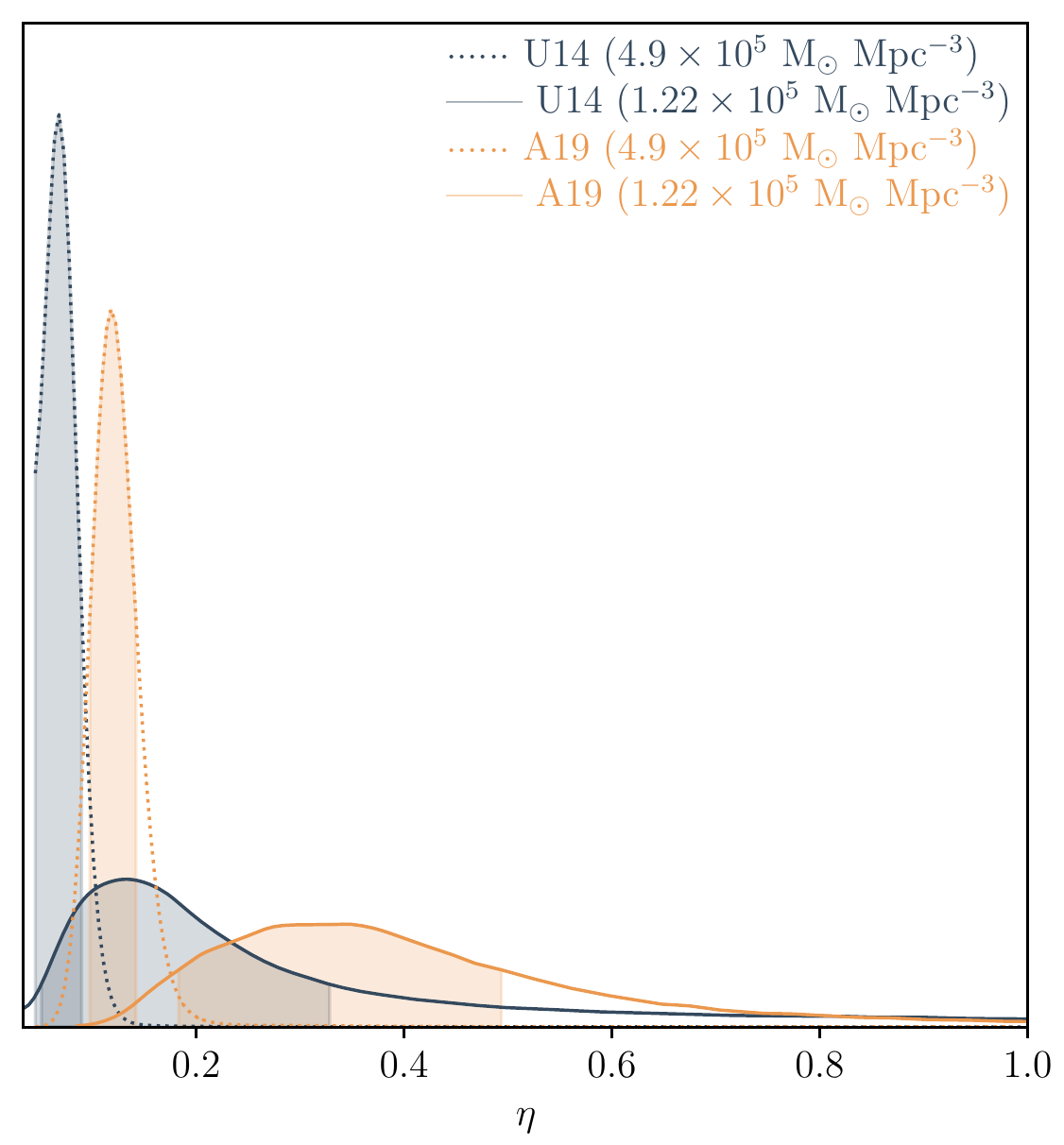}~
	\includegraphics[width=0.45\linewidth]{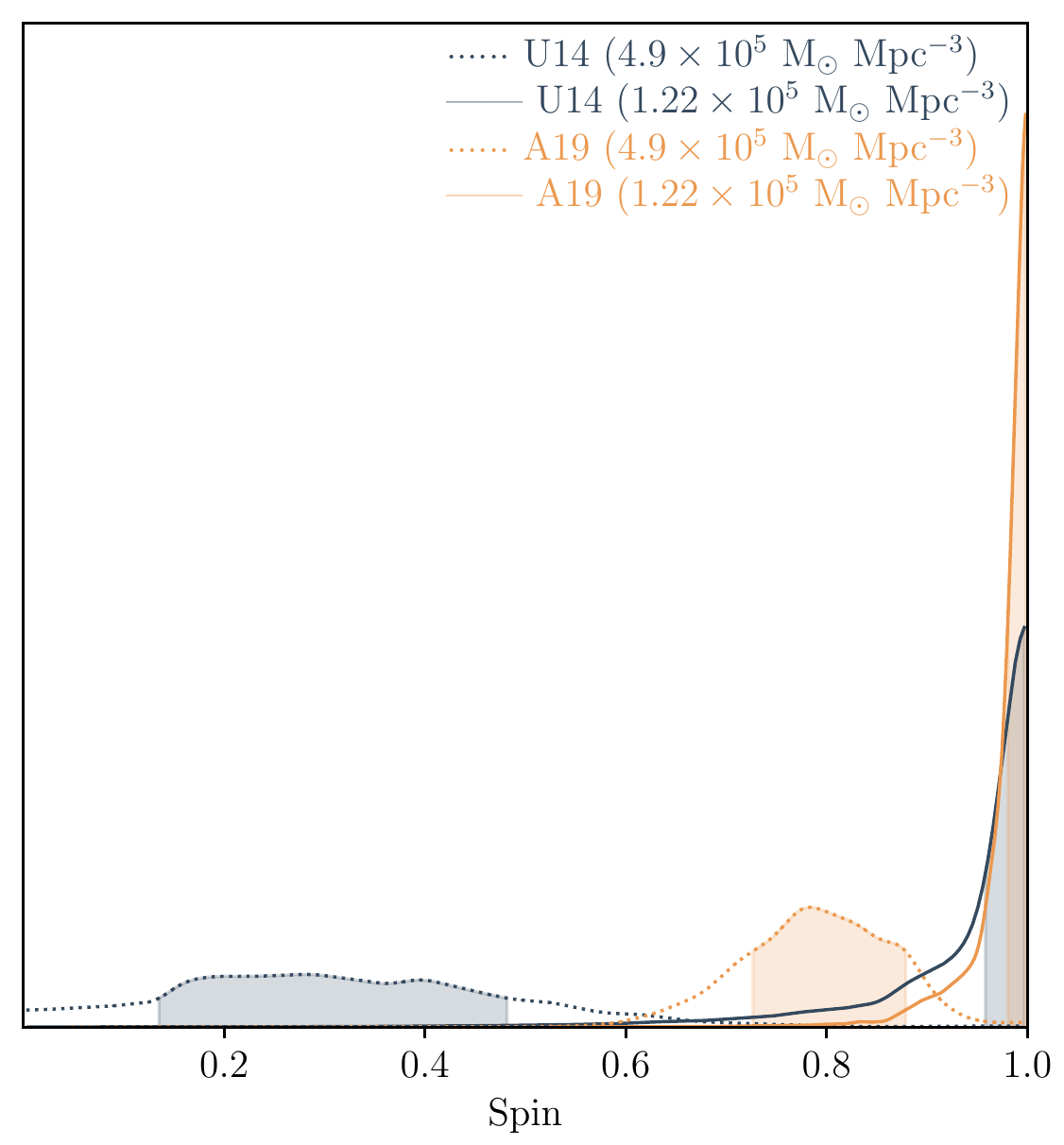}
   	\caption{Probability distributions for radiative efficiency (\textit{left panel}) and black hole spin (\textit{right panel}), derived from the A19 ({\it orange}) and U14 ({\it gray}) XLFs, for two estimates of local SMBH mass density: $1.2 \times 10^5$ M$_\odot$ Mpc$^{-3}$ (\textit{solid lines}), which assumes observed black hole mass measurements are biased toward high masses (\citealp{shankar2019nature}), and $4.9 \times 10^5$ M$_\odot$ Mpc$^{-3}$ (\textit{dotted lines}), the unweighted observed value \citealt{vika2009}). These distributions were computed using MCMC sampling for radiative efficiency $\eta$ with $\eta_{\rm kin} = 0$; \textit{shaded regions} show 1$\sigma$ confidence intervals.
  	The most probable radiative efficiency for A19 (U14) is $\eta=0.34^{+0.15}_{-0.16}~(0.20^{+0.18}_{-0.10})$ for the \citet{shankar2019nature} local black hole mass density.
  	For the higher, unweighted mass estimate, the corresponding numbers are $\eta=0.118^{+0.024}_{-0.020}~(0.068^{+0.022}_{-0.023})$).
  	\textit{Right panel:} Spin distribution calculated from radiative efficiency using the relationship given by \citet{reynolds2012}. The spin values for the higher and lower $\rho_{\rm SMBH}$ estimates are, for A19 (U14), $0.52^{+0.14}_{-0.15}$ 
  	and $0.997^{+0.000}_{-0.03}$ ($0.995^{+0.000}_{-0.0013}$), respectively.}
	\label{fig:simple_radeff} 
\end{figure*}
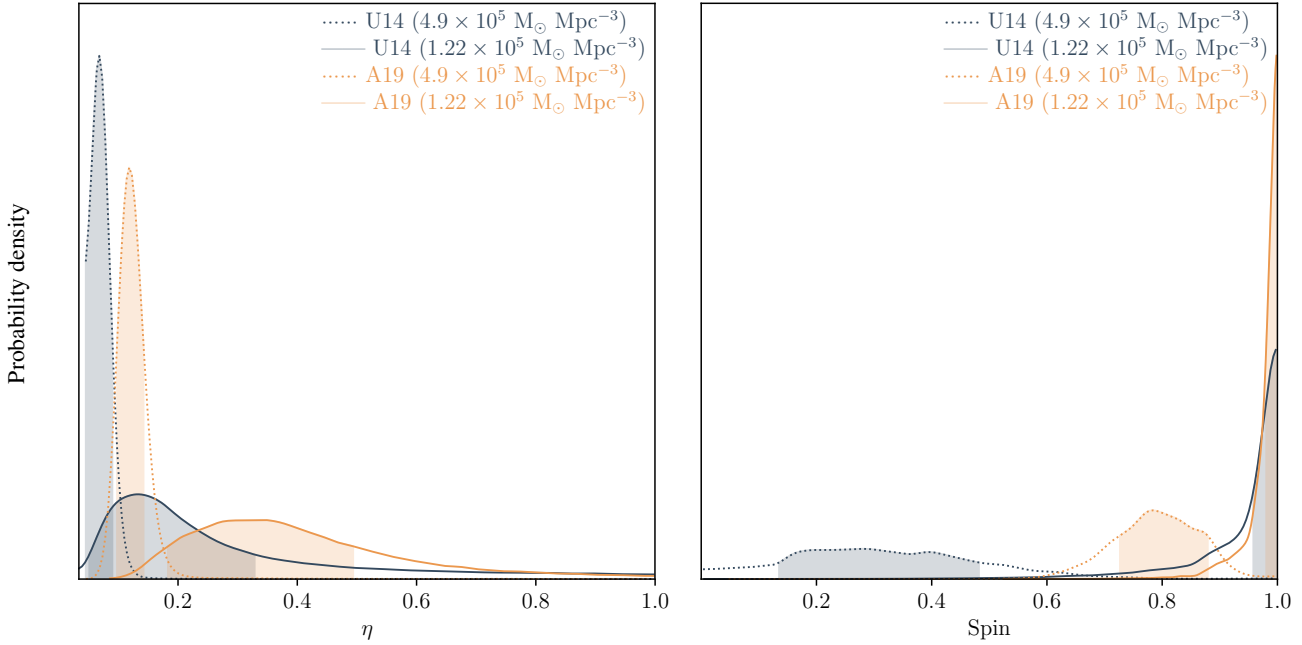 

\begin{figure}[t]
	\centering
	\includegraphics[width=1.\linewidth]{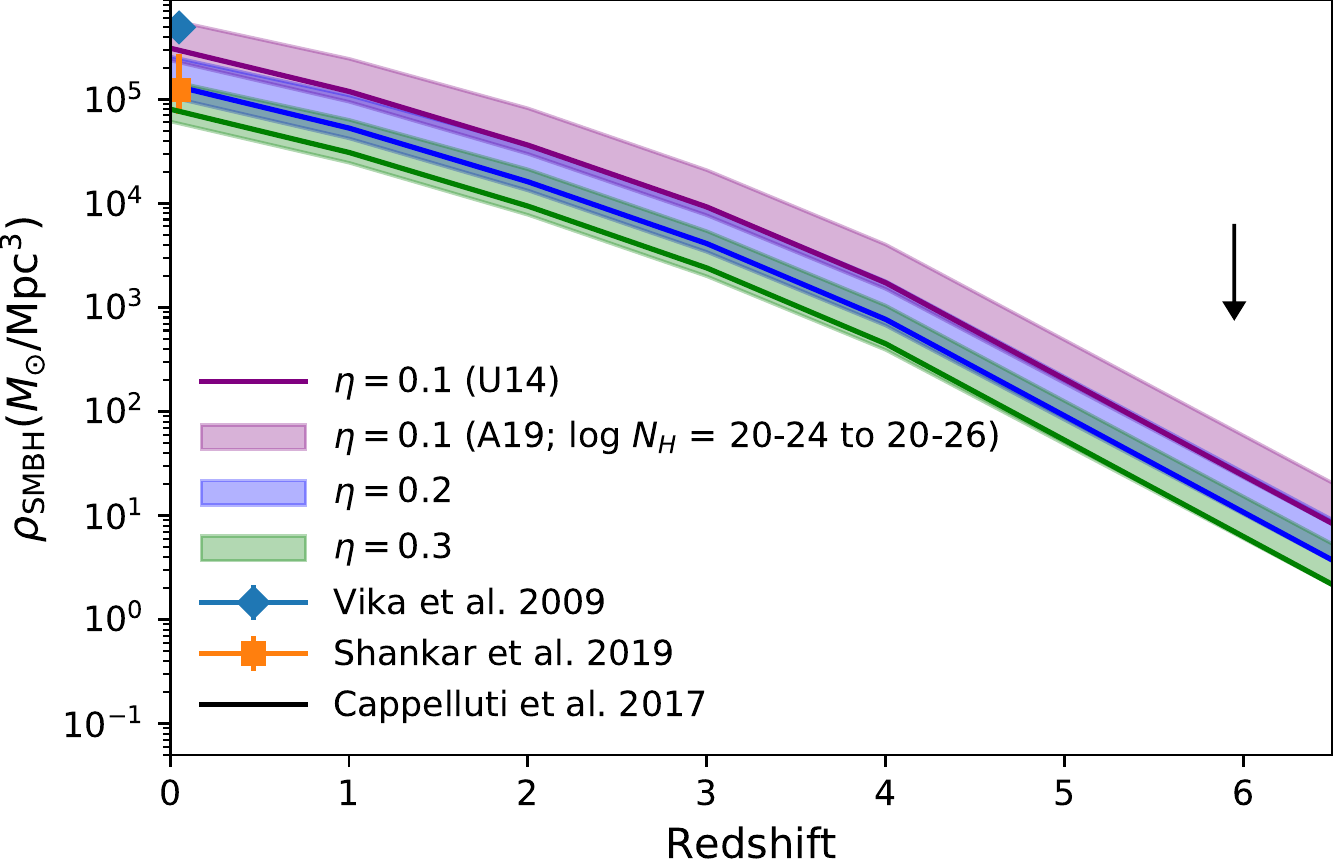}
	\caption{Black hole mass density as a function of redshift for XLFs from A19 ({\it shaded regions}) and 
	U14 ({\it solid lines}), assuming efficiencies $\eta = 0.1$ ({\it purple}), 0.2 ({\it blue}), or 0.3 ({\it green}). The {\it shaded regions} for A19 show the difference in mass density with and without Compton-thick objects: the low end of the shaded region corresponds to Compton-thin objects, with $\log~(N_H / {\rm cm}^{\rm -2}) < 24$, and the upper limit shows mass density including Compton-thick objects. 
	Because A19 accounts for the high space density of Compton-thick objects seen with {\it NuSTAR} and {\it Swift} BAT, it results in a higher $\eta$ than U14. Moreover, because the bias-corrected estimate of local SMBH mass density ({\it orange square}; \citealp{shankar2019nature}) is lower than previous estimates ({\it blue circle}, \citealp{vika2009}), the resulting radiative efficiency is higher still. 
	Our value, $\eta=0.34^{+0.15}_{-0.16}$, implies far more rapidly rotating black holes than previous estimates. The \textit{black arrow} shows the mass density upper limit at high redshift \citep{cappelluti2017}.}
	\label{fig:rho_zs} 
\end{figure} 

In this section we describe our calculation of the AGN contribution to reionization at UV and X-ray wavelengths. 
Then, after converting the A19 XLF into a UV luminosity function, we compare to the the \citet{giallongo2019} and \citet{grazian2020} UV luminosity function for AGN in the redshift range $4<z<6.1$.
We also compare the total ionizing photon contribution from AGN, derived from the A19 model, with the galaxy contribution and overall reionization budget.

\subsubsection{Ultraviolet Emission from AGN}

Using the average intrinsic X-ray spectrum from A19 ($\Gamma = 1.96$, $E_{\rm cutoff} = 200$ keV), we converted the 2$-$10 keV intrinsic AGN luminosity to a luminosity density at 2 keV (erg/s/Hz). The large dispersion in this conversion factor (e.g. Eqn.~9 in \citealp{lusso2010}) causes significant uncertainty in the overall conversion from X-ray to UV wavelengths. Accordingly, we incorporate this uncertainty in the same manner as before, namely:

\begin{multline}\label{eqn:uv_phi}
    \Phi (L_{UV}) = \int \Phi (L_{\rm 2~keV}) \left( \frac{1}{\sigma_{\log L_{\rm UV}} \sqrt{2 \pi}} \right)\\
    \exp{\frac{-(L_{UV}-\langle L_{UV} (L_{\rm 2~keV}) \rangle)^2}{2 \sigma_{\log L_{\rm UV}} ^2}}~dL_{\rm 2~keV} .
\end{multline}

\noindent
To evaluate $\langle L_{UV} (L_{\rm 2~keV}) \rangle$ and $\sigma_{\log L_{\rm UV}}$, we used the inverse of the linear relationship between $\log L_{\rm 2~keV}- \log L_{\rm 2500}$ for X-ray selected quasars \citep{lussorisaliti2016},
\begin{equation}
    \log L_{\rm 2~keV} = 0.638 \log L_{\rm 2500} + 7.074 ,
\end{equation}
to calculate $\langle L_{UV} (L_{\rm 2~keV}) \rangle$. The inverse of $\sigma_{\log L_{\rm 2~keV}} =  0.210$ \citep{lussorisaliti2016} corresponds to $\sigma_{\log L_{\rm UV}} = 0.329$. We then converted the A19 X-ray luminosity function into a UV luminosity function at 1450~\AA\ 
assuming a UV spectral index $\alpha_\nu = 0.61 \pm 0.01$ \citep{lusso2015}.

From this 1450~\AA~UV luminosity function, 
we calculated the number of ionizing photons per comoving volume of the universe,

\begin{equation}
  \dot{n}_{\rm ion} = f_{\rm esc} \xi_{\rm ion} \epsilon_{1450} ,
\end{equation}

\noindent
where $f_{\rm esc}$ is the fraction of UV photons that escape and 
$\xi_{\rm ion}$ is the number of ionizing photons from a quasar with a monochromatic luminosity of 1 erg s$^{-1}$ Hz$^{-1}$ at 1450~\AA\ 
(roughly 10$^{26}$ s$^{-1}/$ [erg$^{-1}$ s$^{-1}$ Hz$^{-1}$]; \citealp{matsuoka2018}).
The total AGN emissivity at 1450~\AA,
$\epsilon_{\rm 1450}$, depends on the luminosity function as follows:
\begin{equation}\label{eqn:epsilon}
\epsilon_{1450} (z) = \int \int \Phi_{A19, 1450~\AA}(z, L_{1450}, \log N _{\rm H}) L_{1450}~dL_{1450}~d\log N _{\rm H} .
\end{equation}

We then evaluated the emissivity at 912~\AA\ 
 assuming $L_{\rm UV} \propto \nu^{-1.7}$ (e.g., \citealt{lusso2015}).
Finally, we calculated the number of ionizing photons per unit co-moving volume for different values of the escape fraction in order to compare to observed values. 

Note that two of the biggest causes of uncertainty in the conversion from X-ray to UV, which we have accounted using Equation~\ref{eqn:uv_phi}, and the unknown escape fraction. \citet{grazian2018} found that for faint AGN (M$_{\rm 1450} = -25.1$ to $-23.3$), the average escape fraction at $3.6<z<4.2$ is 74\%, and extrapolating these results to higher redshift results in a significant contribution from AGN to reionization. However, they 
note that the sample used to calculate this escape fraction is not complete in terms of including both obscured and unobscured AGN, and \citet{cowie2009} found that obscured AGN do not radiate any ionizing photons. Other works have used a more complex method to determine escape fraction. For example, the semi-analytic model of \citet{dayal2020} considered several different cases of escape fraction for star-forming galaxies as well as AGN, with and without redshift and stellar mass dependences. 

If obscuration is mostly due to orientation, we can estimate an empirical escape fraction from the intrinsic ratio of space density of unabsorbed AGN to all AGN. 
Using the A19 luminosity function, 
we obtain the much lower value, $f_{\rm esc}=0.07$. We also considered other relationships between escape fraction and obscuration: for example, \citet{mao2007} assumed $f_{\rm esc} \simeq e^{A_V/1.08}$, and \citet{fricci2017} assumed a constant $\frac{N_H}{A_V}$ ratio, corresponding to $f_{\rm esc} = e^{-(\log N_H - 20)}$.

We compare our results to the \citet{giallongo2019} UV luminosity function, which describes the faint end of the AGN luminosity function ($M_{\rm 1450}$ from $-22.5$ to $-18.5$) at $4 < z < 6.1$, derived from a sample of 32 H band-selected AGN with significant soft X-ray detections with \textit{Chandra}. The H-band flux is rest-frame UV emission ($\lambda < 3000$~\AA) and at $z> 4$, the soft X-ray band (0.5-2 keV) is equivalent to rest-frame 2.5-10 keV. \citet{giallongo2019} corrected this sample for completeness based on the X-ray to H-band flux ratio, which they noted 
was 
not well constrained at the redshifts and luminosities probed. Nevertheless, as UV-selected samples tend to contain mainly unabsorbed AGN, a completeness correction might be able to recover at least the more heavily obscured Compton-thin objects ($\log N_H = 22-24$) which are detectable in the $2-10$ keV band (U14); therefore, we compare our results with the total emissivity found by \citet{giallongo2019}. 

\subsubsection{X-Ray Emission from AGN}\label{sec:xray_method}

The X-ray spectral parameters of AGN, as constrained by extensive available data, are discussed in detail by A19 and \citet{ananna2020}.
Integrating these spectra over energy, $N_{\rm H}$, and luminosity yields the number of ionizing photon per unit volume as a function of redshift:

\begin{multline}
  \dot{n}_{\rm X-ray} (z) = \int^{L_{\rm max}}_{L_{\rm min}} \int^{\log N_{\rm H, max}}_{\log N_{\rm H, min}} \int^{E_{\rm max}}_{E_{\rm min}} \Phi_{\rm A19} (z, L_{\rm X}, N_{\rm H})
  \\
  \frac{dN}{dE}(E, L_{\rm X}, \log N_{\rm H})~dE~d \log N_{\rm H}~d L_{\rm X} ,
\end{multline}

\noindent
where $E$ is the photon energy 
and $\frac{dN}{dE}(E, L_{\rm X}, \log N_{\rm H})$ is the photon spectrum of each AGN, i.e., the number of photons produced by an AGN of intrinsic luminosity L$_X$ as a function of energy and absorbing column $\log N_{\rm H}$. 
We use the best-fit X-ray spectra described in \S~5.2 and \S~6 of A19, which have the following parameters taken from the \textit{Swift}-BAT 70-month sample \citep{claudio2017bat}: $\Gamma = 1.96 \pm 0.1$, $E_{\rm cutoff} = 200 \pm 29$ keV, $R_{\rm obscured} = 0.37 \pm 0.1$, $R_{\rm unobscured} = 0.83 \pm 0.1$ and $F_{\rm scatt} = 1\%$. We use the updated torus model \textsc{borus02} (\citealp{mislav2018}; as in \citealp{ananna2020}) instead of \textsc{bntorus} (\citealp{bntorus2011}; as in A19). 
When repeating the calculation for the U14 XLF, we used the absorption function described in Table~2 of that work, where the total space density of Compton-thick objects is assumed to be equal to the total number density of Compton-thin objects.
We integrate the XLF from $N_{\rm H,max} = 10^{20}$ to $10^{26}$~atoms cm$^{-2}$, and from $E=$0.5 to 500~keV, where 
the spectral shape is relatively well constrained (e.g., \citealp{claudio2017bat}).




\subsubsection{Total Ultraviolet Emission and Contribution from Galaxies}
The 
history of reionization is still uncertain
because over the last two decades, different probes have reported vastly different estimates of the dominant redshift of reionization. Measurements of the Thomson optical 
depth from polarization of the cosmic microwave background have placed the epoch of reionization variously at $z= 20^{+11}_{-9}$ (\citealp{spergel2003}, early WMAP result);
$z= 10.6 \pm 1.1$ (\citealp{bennet2013}, later WMAP result);
and $z\sim 6.93-7.8$ (Planck results from \citealp{planck2019,efs2019,mason2019}). 
These lower redshift estimates allow for a significant contribution from AGN as well as stars in galaxies.

\citet{bouwens2015b} constructed a two-parameter model of the evolution of total cosmic ionizing emissivity, $\dot{N}_{\rm ion} (z)$, using observed values at $z> 6$ by Planck and WMAP. This model and the UV luminosity density of galaxies according to the \citet{bouwens2015} galaxy luminosity function suggest that star-forming galaxies provide the ionizing photons required to reionize the universe. Their analysis of several AGN optical and UV LFs \citep{willott2009,willott2010,mcgreer2013} indicate that quasar emissivity is too low to be a significant contributor to reionization, although at the time faint-end of the quasar luminosity function was rather uncertain. Recently, the faint-end has been updated by \citet{giallongo2019} and \citet{grazian2020}, so we compare A19 XLF with these new high redshift quasar number densities as well as with the \citet{bouwens2015b} results.

\citet{mason2019} derived a non-parametric form of total $\dot{N}_{\rm ion} (z)$ by combining constrains from optical depth, dark gap statistics, Ly$\alpha$ damping wings of quasars and ratio of Ly$\alpha$ emitters over Lyman break galaxies. They also provided estimates of the galaxy contribution using the UV LF from \citet{mason2015}, assuming a constant escape fraction of 20\% for all galaxies. We included both the total $\dot{N}_{\rm ion} (z)$ and galaxy contribution estimates from this work in our analysis.

\citet{dayal2020} used a semi-analytic model to reproduce galaxies and black holes, and explore a wide-range of combinations of escape fractions. For the fiducial model, they assumed that a star-forming galaxies with a SMBH will have the same escape fraction as the AGN (determined using ratio of AGN with $\log N_H$ $< 22$ to all AGN), while all other galaxies will have an escape fraction of zero. The galaxy ionization contribution predicted by this model 
is one of the lowest estimates reported by \citet{dayal2020}, only slightly higher than 
a model with galaxy $\langle f_{\rm esc} \rangle = 1.1 (\frac{1+z}{7})^{3.8}$. We compare both these models 
to our results in \S~\ref{sec:results}.

\section{Results}\label{sec:results}

\subsection{Spin and Radiative Efficiency of SMBH}

For a constant radiative efficiency, i.e., independent of redshift, mass or luminosity, the MCMC chain converges rapidly (in $<100$ steps). Table~\ref{tab:A18_mcmc} gives the resulting efficiencies under several different assumptions.
Using the local SMBH mass density 
from \citet{vika2009} and the A19 XLF, we obtain $\eta = 0.118^{+0.024}_{-0.020}$; the full probability distribution is shown in the left panel of Figure~\ref{fig:simple_radeff} (\textit{orange dotted line}).
This is 
similar to previous values (e.g., \citealp{yutremaine2002}, \citealp{sijacki2015}), 
and roughly double what we find using the U14 luminosity function (which we now know does not account for all the Compton-thick AGN), $\eta = 0.068^{+0.022}_{-0.023}$ (\textit{black dotted line} in Fig.~\ref{fig:simple_radeff}), which is close to the lowest theoretically allowed value. The A19 XLF results in a higher efficiency than the U14 XLF because it produces more light, roughly half from Compton-thick AGN. 

For the bias-corrected SMBH mass density of \citet{shankar2019nature}, which is plausibly closer to the true value, we obtain a significantly higher radiative efficiency, $\eta = 0.34^{+0.15}_{-0.16}$ ($0.20^{+0.18}_{-0.10}$ for U14), as shown by the solid lines in Figure~\ref{fig:simple_radeff}. 
Such a high efficiency implies most black holes are rotating rapidly. The distributions in spin 
are shown in the right panel of Figure~\ref{fig:simple_radeff}. 
Note that the spins for A19 and U14 overlap almost completely, even though the efficiencies differ by a factor of two, due to the non-linear mapping from $\eta$ to spin (\citealp{reynolds2012}). 
For the lower local SMBH mass density, 
the derived spin value is
$0.998^{+0.000}_{-0.018}$ (A19).
Instead using the higher value for SMBH density, 
the spin would be slightly lower $(0.783^{+0.096}_{-0.056})$. 
We discuss these results in \S~\ref{sec:conclusion}.

In all cases, $\eta$ is significantly lower for $\eta_{\rm kin} = 0.15$ compared to $\eta_{\rm kin} = 0$ (see Table~\ref{tab:A18_mcmc}). This is expected because some of the accretion energy goes into outflows rather than radiation. However, spin in higher for $\eta_{\rm kin} = 0.15$ as the total spin is dependent upon the sum of radiative and kinetic efficiencies.
In Figure~\ref{fig:rho_zs} we show the evolution of mass densities of SMBH for several values of constant 
radiative efficiency, for both the A19 and U14 XLFs ({\it solid} and {\it dotted lines}, respectively). 
This directly shows how the accreted mass depends on the space density of Compton-thick objects ({\it solid versus dotted lines}) and on the local SMBH mass density assumed (\textit{blue versus orange data points}). Specifically, adopting the bias-corrected local mass density ({\it orange square}), which is lower by a factor of $\sim4$ than the \citet{vika2009} value, increases the derived efficiency by the same factor. 



\subsection{Contribution to Cosmic Reionization}

\begin{table*}
   \centering
   \caption{Radiative Efficiency\tablenotemark{(a)} and Spin\tablenotemark{(b)}}
   \label{tab:A18_mcmc}
   \begin{tabular}{cccc}
       \hline
		 XLF & $\eta_{\rm kin}$ & $\eta$\tablenotemark{(c)} & Spin \\ 
		\hline
		 \textbf{A19\tablenotemark{(d)}} & 0 & $0.34^{+0.15}_{-0.16}~(0.118^{+0.024}_{-0.020})$ & $0.998^{+0.000}_{-0.018}~(0.783^{+0.096}_{-0.056})$ \\
		 & 0.15 & $0.30^{+0.12}_{-0.15}~(0.100^{+0.021}_{-0.019})$ & $0.9995^{+0.0}_{-0.0012}~(0.9832^{+0.0039}_{-0.0034})$ \\ 
		 \hline
		 \textbf{U14} & 0 & $0.20^{+0.18}_{-0.10}~(0.068^{+0.022}_{-0.023})$ & $0.997^{+0.000}_{-0.039}~(0.28^{+0.20}_{-0.15})$ \\
		 & 0.15 & $0.178^{+0.162}_{-0.098}~(0.059^{+0.021}_{-0.023})$ & $0.99961^{+0.0}_{-0.00081}~(0.9632^{+0.0056}_{-0.0037})$ \\
		\hline
    \end{tabular}
    \tablenotetext{(a)} {~~~Calculated for different XLFs and values of $\rho_{\rm SMBH}$.}
    \tablenotetext{(b)} {~~~Converted from radiative efficiency using Figure~6 in \citep{reynolds2012}.}
    \tablenotetext{(c)} {~~~For local SMBH mass densities $\rho_{\rm SMBH} = 1.2 \times 10^5 M_\odot$ Mpc$^{-3}$ ($\rho_{\rm SMBH} = 4.5 \times 10^5 M_\odot$ Mpc$^{-3}$).}
    \tablenotetext{(d)} {~~~Presently the only XLF that fits all available data, including recent {\it NuSTAR}, {\it Swift} BAT and {\it Chandra} Deep Field South results.}
\end{table*}

\begin{figure}[t]
	\centering
	\includegraphics[width=1.\linewidth]{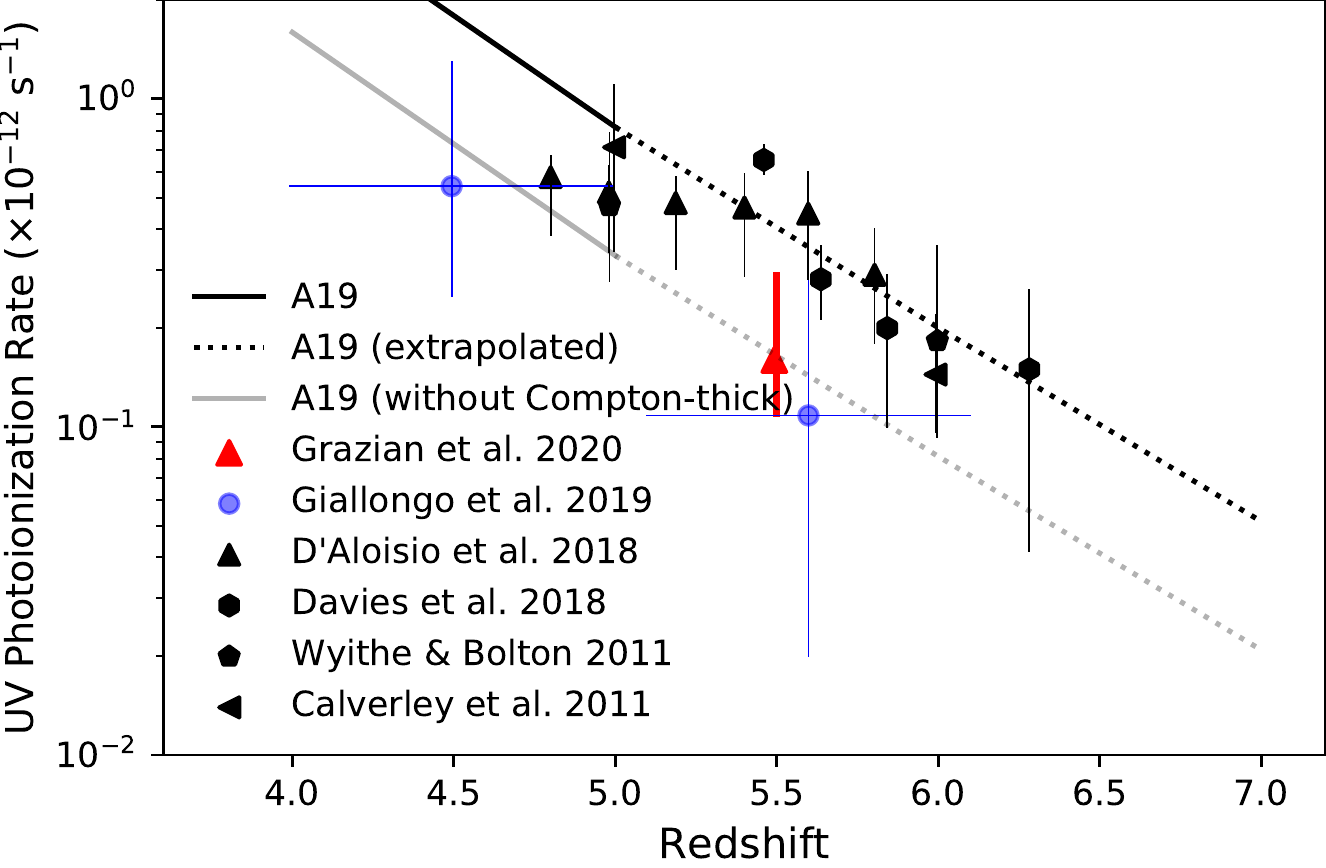}
	\caption{ Cosmic photoionization rate 
	produced by AGNs as a function of redshift. {\it Black} and \textit{blue data points} are from Figure~6 of \citet{giallongo2019}. The \textit{blue} point represent their
	UV luminosity function from a sample of unobscured quasars, and \textit{red data point} represents an updated estimate by \citet{grazian2020}.
	The UV emission derived from the A19 XLF (\S~\ref{ssec:UVconv}) agrees well with these data: \textit{grey lines} include only objects with $\log N_{\rm H} < 24$ (corresponding to the unobscured quasars sampled by \citealp{giallongo2019}) and \textit{bold lines} include the contribution of Compton-thick AGN. 
	The escape fraction for all AGN was assumed to be unity.}
	\label{fig:giallongo} 
\end{figure}

As an initial check on our XLF to UV LF conversion, we 
compare our derived photoionization rate to the one from \citet{giallongo2019}, 
after multiplying each by a factor of 1.2 to account for the contribution of recombination radiation to the ionizing background \citep{daloisio2018, FJ2009}. 
We find that the summed space densities of unabsorbed and Compton-thin objects from A19 agree well with \citet{giallongo2019} and \citet[the latter provides an updated estimate of the former]{grazian2020}. Therefore, we also compute an upper limit to reionization from AGN, by considering the high escape fraction suggested by \citet{grazian2018} for both unabsorbed and Compton-thin AGN ($\log N _{\rm H} < 24$) at all redshifts and luminosity. Figure~\ref{fig:giallongo} shows that the A19-based UV photoionization rate for AGN with $\log N_{\rm H} < 24$ ({\it grey lines}) agrees well with the \citet{giallongo2019} and \citet{grazian2020} UV LFs (\textit{blue and red points}, respectively), both where AGN space densities are well constrained ($z<5$) and extrapolated to higher redshifts.
When we add radiation from Compton-thick AGN ({\it black lines}), our results agree with the observed data (\textit{black points}) from \citet{calverley2011}, \citet{wyithe2011}, \citet{daloisio2018} and \citet{davies2018}.

\begin{figure*}[t]
	\centering
	\includegraphics[width=1.\linewidth]{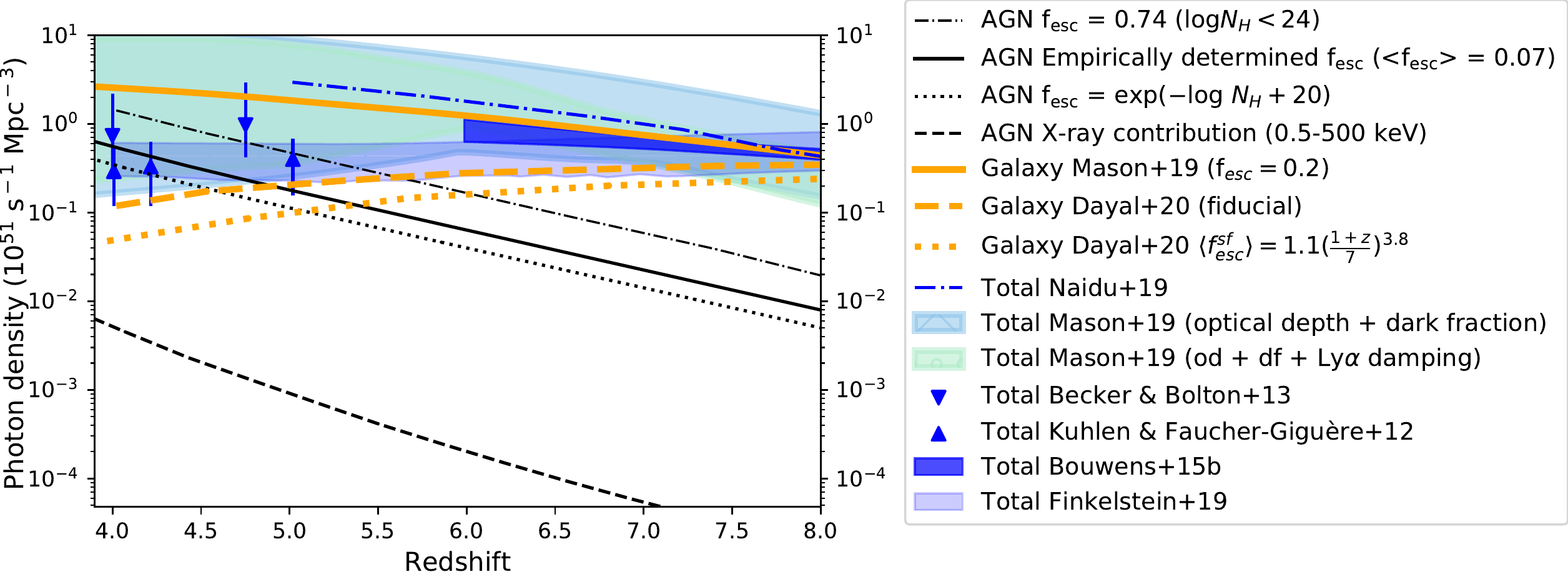}
	\caption{Ionizing photon densities for AGN (\textit{black lines}) and galaxies (\citealp[{\it orange dashed and dotted lines}]{dayal2020}; 
	\citealp[{\it orange solid line}]{mason2019}), and total ionizing contribution (\citealp[{\it blue triangles}]{kuhlen2012}; 
	\citealp[{\it blue inverted triangles}]{bb2013};
	\citealp[{\it sky blue and teal shaded regions}]{mason2019};
	\citealp[{\it dark blue shaded region}]{bouwens2015b};
	\citealp[{\it light blue shaded region}]{finklestein2019}).
	 UV light from AGN, derived from the A19 XLF ({\it black lines}), contributes substantially at moderate redshifts, while galaxy light \citep[\textit{ orange dashed and dotted lines}]{dayal2020} dominates above $z\sim6$.
	 For an escape fraction $f_{\rm esc}=0.07$, corresponding to the average ratio of unobscured to total AGN, 
	 the AGN contribution at $z\sim6$, integrated over all $N_{\rm H}$, is {23\% of the lowest galaxy contribution shown by \textit{dashed orange line}}. For an escape fraction as high as $f_{\rm esc}=0.74$, estimated for a sample of unobscured AGN \citet{grazian2018} and thus an upper limit to the true value, the UV emission from AGN with $\log N _{\rm H} < 24$ at $z>6$ is {59.8\% of galaxy contribution shown by \textit{dashed orange line}, or 37.4\% of the total contribution}.
	 At all redshifts, the AGN photon density at UV wavelengths (\textit{black solid} and {\it dot-dash lines}) is at least 20 times higher than at X-ray wavelengths (\textit{black dashed line}). }
	\label{fig:photon_count} 
\end{figure*}

In Figure~\ref{fig:photon_count}, we compare the ionizing photon densities for models of AGN and galaxy emission, as a function of redshift, to the total ionizing photon density inferred from hydrogen absorption along the line of sight. 
As can be seen from the figure, for the empirically determined escape fraction, \textbf{$f_{\rm esc}=0.07$,} the A19-derived ionizing radiation from AGN
 produces 1.7-23.1\% of the 1$\sigma$ upper \citep{mason2019} and lower \citep{finklestein2019} limits of total ionizing budget required at $z\sim6$, respectively. The contribution decreases at higher redshifts, where the lowest galaxy contribution \citep{dayal2020} is 10 to 100 times that of AGN.
 Using a higher escape fraction, $f_{\rm esc} = 0.74$, derived by \citet{grazian2018} for faint UV-selected AGN at $z\sim 3.6-4.2$ (integrating the LF for $\log N_{\rm H} \leq 24$), star formation and AGN contribute roughly equally at $z \sim 6$. However, such high escape fractions are likely unrealistic for Compton-thin AGN. 
 We discuss these results 
 in \S~\ref{sec:conclusion}.
 
 We calculated the X-ray contribution from AGN using the method described in \S~\ref{sec:xray_method}. As expected from the shape of AGN spectra (e.g., \citealp{harrison2014}), the number density of ionizing photons at X-ray wavelengths is about an order of magnitude lower than the UV ionizing photon density at $z< 3$, and even lower at high redshifts. Therefore, the AGN X-ray contribution to ionizing photon density is negligible compared to the AGN UV contribution, and less than $0.1$\% of galaxy contribution at $z>6$.
 

\section{Summary and Discussion}\label{sec:conclusion}

We have quantified the growth of SMBH and their ionizing radiation as a function of redshift, based on X-ray samples that are relatively unbiased because high energy X-ray photons can penetrate heavy obscuration. Specifically, we used the recent A19 X-ray luminosity function, which fits all current observed X-ray constraints on AGN populations, converting it to a bolometric luminosity function {by taking the dispersion in X-ray to bolometric conversion into account}. Integrating over all energies and redshifts, and comparing this total radiation to the local mass density of SMBH, we calculated the average efficiency with which SMBH convert the gravitational potential energy of infalling matter into light.
We also investigated the AGN contribution to reionization. 

The derived radiative efficiency depends crucially on the estimated value of local SMBH mass density, which is still quite uncertain. \citet{vika2009} and \citet{li2011} reported a local SMBH mass density of $\sim 4.5-5 \times 10^5$ M$_\odot$ ${\rm Mpc}^{-3}$; {using this value, we obtained an average efficiency $\eta = 0.1-0.12$ (for $\eta_{\rm kin} = 0.15-0$), similar to previous estimates using similar local densities.
However, \citet{shankar2019nature} reported a mass density that is $\sim 4$ times lower, $1.2 \times 10^5$ M$_\odot$ ${\rm Mpc}^{-3}$, after correcting for the observational bias toward larger SMBH. 
This improved value leads to a radiative efficiency, $\eta = 0.34^{+0.15}_{-0.16}$.} The high efficiency {results in part} because A19 XLF includes more Compton-thick AGN than other recent XLFs (e.g., U14, \citealp{aird2015}), as required by new {\it NuSTAR} and {\it Swift}-BAT results \citep{civano2015, lansbury2017,ricci2017ulirg,lanzuisi2018,masini2018,marchesi2018,marchesi2019}, roughly doubling the total radiation. 

According to some theoretical estimates, the maximum radiative efficiency could be as low as $\eta = 0.3$ \citep{thorne1974}, 
lower than our peak value (for the \citealp{shankar2019nature} lower mass density) 
but well within the uncertainties. 
Like A19, a recent mid-infrared study found a substantial population of heavily obscured objects that were previously missed by X-ray and optical surveys, implying a higher radiative efficiency than previously estimated \citep{lacy2020}. Constraining the number density of Compton-thick objects is a work in progress, but the new hard X-ray survey data, on which A19 was based, are inconsistent with the lower Compton-thick fraction in the XLFs such as \citet{ueda2014} and \citet{aird2015}. 
Perhaps more uncertain is the correct value of the local mass density, which has a big effect on the derived radiative efficiency and which will improve as the statistics of black hole masses improve.


Because the relation of spin to efficiency is highly nonlinear \citep{reynolds2012}, the updated XLF and mass density yield an average spin that is much higher than previous estimates, as shown in Figure~\ref{fig:simple_radeff}. Our results as well as U14 suggests that most AGN could be spinning close to the maximal value. 

Radiative efficiency may well be more complicated than the uniform value assumed here. In particular, it may depend on accretion rate, black hole mass, Eddington ratio, and/or redshift. 
\citet{li2012} calculated redshift- and mass-dependent radiative efficiency and found that massive black holes ($M_{\rm BH} \geq 10^{8.5} M_{\odot}$) spun down from high radiative efficiencies of $\eta \sim 0.3-0.4$ at z $\geq$ 1 to an order of magnitude lower values by z $\sim 0$, whereas less massive black holes maintained efficiency at $\eta \leq 0.2$. 
We are currently calculating a new local black hole mass function and Eddington ratio distribution function (for a separate work), after which we will undertake a full consideration of the possible dependency of radiative efficiency on these variables. 

{Examining the AGN contribution to cosmic reionization by extrapolating the A19 XLF beyond $z = 5$, we conclude that galaxies are still the dominant contributor to reionization for realistic escape fractions, 
in agreement with \citet{shankarmathur2007} and \citet{robertson2015}.
However, if we assume very high escape fraction ($f_{\rm esc} = 0.74$) and approximately the lower limit for galaxy contribution (\textit{dashed orange line} in Figure~\ref{fig:photon_count}) for unobscured and absorbed Compton-thin AGN ($\log~N_H < 24$), 
AGN could provide as much as 37\% of the total UV radiation at $z\sim 6$.} 
However, the escape fraction is likely to be much lower since the 
\citealp{grazian2018} sample was dominated by unobscured AGN).
X-ray surveys indicate that in the local universe 70\% of all AGN are obscured \citep{ricci2015}, and this high fraction appears to increase with redshift (\citealp{treister2006}, U14, A19, \citealp{vito2019}). It is likely that studies based on optical- and UV-selected AGN samples are biased toward low covering factors and thus artificially high values for the escape fraction.
The A19 XLF indicates that half of total AGN light comes from Compton-thick objects, suggesting that the escape fraction is low.  {For our empirically determined value, $f_{\rm esc}=0.07$, the AGN contribution to reionization is less than a quarter of the total ionizing photon density at $z \geq 6$.}

The foundation on which this works stands --- namely, the A19 space density of AGN as a function of luminosity, obscuring column density, and redshift --- is fully general and accurately explains the vast array of X-ray data now available. 
In comparison, our assumptions that escape fraction and radiative efficiency are constant with redshift and AGN properties are much simplified. Still, our basic conclusions are not affected: allowing more complicated parameter dependencies would not obviously decrease radiative efficiency, and although the fraction of ionizing photons that escape from AGN and galaxies is uncertain (e.g., \citealp{lusso2015,grazian2018,giallongo2019,dayal2020}),
we have explored the full range of values, so a more complicated approach would not enhance the AGN contribution to reionization.
	
The AGN contribution to reionization also depends on other parameters that are poorly constrained, such as the shape of the quasar ionizing continuum (e.g., see discussion in \citealp{lusso2015}) and the uncertain space density of AGN at high redshifts. 
So it is remarkable that the photoionization rates computed from the A19 XLF (\S~\ref{ssec:UVconv}) and from \citet{giallongo2019} and \citet{grazian2020} agree when we assume the same escape fraction of 1 for both works for $\log N_H < 24$ (Fig.~\ref{fig:giallongo}). In contrast, \citet{fricci2017} found that \citet{ueda2014} implied lower space densities than \citet{giallongo2015} even after including Compton-thick objects (Figure~1 of that work). Note that comparing Figure~4 of \citet{giallongo2015}, Figure~4 of \citet{giallongo2019} and Figure~5 of \citet{grazian2020}, the latter papers also report lower space densities of AGN at high redshifts than \citet{giallongo2015}. Our work agrees with the more up-to-date results.

In the end, the key variable proves to be the AGN number density rather than spectral shape or escape fraction. The A19 XLF is based on the most complete, unbiased AGN samples available, critically including high-energy X-rays ($E>$10~keV), uncovering a large population of Compton-thick AGN that is missed in optical and UV surveys. This comprehensive census of SMBH growth, which is integral to understanding galaxy formation and cosmic evolution, points to two important conclusions. 
First, accretion dominates over mergers for black hole growth, as the former spins up black holes (adding orbital angular momentum of accreting particles) while the latter should more often reduce spin (because the angular momenta of merging black holes are randomly oriented).
Second, AGN are relatively unimportant in reionizing the universe at $z>6$, compared to galaxies. Interestingly, their photon output catches up with galaxies toward cosmic noon ($z \simeq 2$), as SMBH and their peak emissivity grow.
A more detailed study of the evolution of SMBHs, in both mass and emissivity, may reveal an imprint from the evolving impact on galaxies and the IGM.

\noindent
{We wish to thank the referee for insightful comments that helped improve the quality of the analysis}. TA wishes to thank Roberto Gilli for a close reading of the paper and thoughtful comments. This material is based upon work supported by the National Science Foundation under Grant No. AST-1715512, NASA under ADAP Grant No. 80NSSC18K0418, and Yale University. ET acknowledges support from FONDECYT Regular 1160999 and 1190818, ANID PIA ACT172033 and Basal-CATA AFB170002. RCH acknowledges support from the National Science Foundation under grant 1554584, and NASA under grant 80NSSC19K0580. FS acknowledges partial support from a Leverhulme Trust Research Fellowship. CR acknowledges FONDECYT Iniciacion grant 11190831.
TA wishes to thank her family members, M. A. Quayum, Shamim
Ara Begum, Mehrab Bakhtiar, Arnita Tasnim Ankur and Raysa Tasnim Orin for their
support.

\textit{Software:} {\tt numpy} \citep{numpy2011}, {\tt Astropy} \citep{astropy2018}, {\tt Emcee} \citep{emcee},
{\tt Matplotlib} \citep{matplotlib}, {\tt Topcat} \citep{taylor2005}, \textsc{xspec} and \textsc{pyxpsec} \citep{xspec}, ChainConsumer and {\tt Vegas} \citep{vegas}.

	
\end{document}